\documentclass[11pt,a4paper]{article}
\usepackage{jheppub}

\usepackage{epsfig}
\usepackage{amsmath}
\usepackage{amssymb}
\usepackage{amsbsy}
\usepackage{enumerate}
\usepackage{bbm}
\usepackage{url}
\usepackage[hang,sf]{subfigure}
\def\beq{\begin{equation}}
\def\beqn{\begin{eqnarray}}
\def\eeq{\end{equation}}
\def\eeqn{\end{eqnarray}}

\def\({\left(} 
\def\){\right)}

\newdimen\figwidth
\figwidth=\textwidth

\newcommand\POWHEG{{\tt POWHEG}}

\newcommand\PYTHIA{{\tt PYTHIA}}
\newcommand\SHERPA{{\tt SHERPA}}
\newcommand\Blackhat{{\tt BlackHat}}
\newcommand\MCFM{{\tt MCFM}}

\newcommand\BOX{{\tt POWHEG BOX}}
\newcommand\MCatNLO{{\tt MC@NLO}}

\newcommand\sss{\mathchoice%
{\displaystyle}%
{\scriptstyle}%
{\scriptscriptstyle}%
{\scriptscriptstyle}%
}
\newcommand\nplus{\oplus}
\newcommand\nminus{\ominus}
\newcommand\splus{{\sss \nplus}}
\newcommand\sminus{{\sss \nminus}}

\newdimen\hbigcirc
\newdimen\wbigcirc

\newcommand\ep{\epsilon}

\newcommand\as{\alpha_{\sss\rm S}}

\newcommand\pt{p_{\sss\rm T}}

\newcommand\kt{k_{\sss\rm T}}

\newcommand\PSn{\Phi_n}

\newcommand\mydot{\!\cdot\!}

\newcount\minutes 
\newcount\scratch 
 
\def\timestamp{
\scratch=\time 
\divide\scratch by 60 
\edef\hours{\the\scratch} 
\multiply\scratch by 60 
\minutes=\time 
\advance\minutes by -\scratch 

\today\ --$\,$\hours:\null 
\ifnum\minutes< 10 0\fi 
\the\minutes
} 

\def\nn{\nonumber}

\title{NLO corrections merged with parton showers for $Z+2$~jets
  production using the {\tt\bf POWHEG} method}
\author{Emanuele Re}
\affiliation{Institute for Particle Physics Phenomenology, Department of Physics\\
  University of Durham, Durham, DH1 3LE, UK}
\emailAdd{emanuele.re@durham.ac.uk}
\abstract{We present results for the QCD production of $Z/\gamma +2$
  jets matched with parton showers using the \POWHEG{} method. Some
  technicalities relevant for the merging of NLO corrections for this
  process with parton showers are discussed, and results for typical
  distributions are shown, in presence of different sets of cuts. A
  comparison with ATLAS data is also presented, and good agreement is
  found.}
\begin{document}
\maketitle














\section{Introduction}
\label{sec:introduction}
The process of vector boson production in association with jets is a
very important Standard Model process at hadron colliders.  $W$-boson
production in association with jets is a typical background for many
new-Physics searches, and it is also interesting as a process in
itself to study jets at hadron colliders. The production of a
$Z$-boson in association with jets present similar features, since for
example when the $Z$ decays in a neutrino pair one can have a
``missing energy + jets'' signature~\cite{Bern:2011pa}, and it is also a reducible
background for $ZH$ production, similarly to the $W+2$ jets case for
$WH$ production.  Moreover, the production of a $Z$-boson in
association with 2 jets is also one of the backgrounds to Higgs-boson
production via vector-boson fusion (VBF), for example when the Higgs
decays in a $\tau^+ \tau^-$ pair. In particular, the signal-background
separation between $Hjj$ via VBF
and $Zjj$ relies on the very different kinematical properties of the
jets produced in association with the heavy
boson~\cite{Rainwater:1998kj,Plehn:1999xi}. 


For the above reasons, it is important to reach a high-level of
accuracy in the predictions for the production of $Z+2$ jets at hadron
colliders.  The full Next-to-Leading Order (NLO) corrections for this
process have been computed in
refs.~\cite{Campbell:2002tg,Campbell:2003hd} for the QCD production,
and in ref.~\cite{Oleari:2003tc} for the electroweak production.
Nowadays, one of the methods to improve upon existing NLO computations
is to merge them with parton showers. This has been realized with the
\MCatNLO{}~\cite{Frixione:2002ik,Frixione:2003ei} and the
\POWHEG{}~\cite{Nason:2004rx,Frixione:2007vw} methods for a number of
processes of increasing
complexity~\cite{Alioli:2010qp,Alioli:2010xa,Kardos:2011qa,Alioli:2011as,
  D'Errico:2011sd,Hoeche:2011fd}. In particular, in the recent past,
$W+2$ and $W+3$ jets have been implemented with the \MCatNLO{}
method~\cite{Frederix:2011ig,Hoeche:2012ft}, and $H+2$ jets using
\POWHEG{}~\cite{Campbell:2012am}.  Given the role played by jet
activity in many searches where $Z+2$ jets is a background, it is
important to achieve the same level of accuracy for this process as
well.

In this paper NLO corrections for the QCD production of $Z/\gamma+2$
jets merged with parton showers according to the \POWHEG{} method are
presented for the first time. Effects due to photon exchange and all
the spin correlations of the decay products of the $Z/\gamma$
intermediate state have been fully taken into account in the
implementation, although in the following we will refer to the process
simply as ``$Z+2$ jets production''. We will concentrate here mainly
on the implementation procedure and on the discussion of the results
we obtained, in presence of different cuts, leaving a broader
phenomenological study for future work, although a comparison with the
first available LHC data is presented.

\section{Method and details of the implementation}
\label{sec:method}
To simulate with NLO matched with parton shower (NLO+PS) accuracy the
$Z+2$ jets process,
the \BOX{} package~\cite{Alioli:2010xd} has been used. The \BOX{} is a
program that automates all the steps described in
ref.~\cite{Frixione:2007vw}, turning a NLO calculation into a
\POWHEG{} simulation. The details of how the program works have been
largely described in ref.~\cite{Alioli:2010xd}, and therefore will not
be repeated in this paper. In this section, we first summarize how the
inputs needed by the package to work were obtained, and we then
describe some technical details relevant for this implementation.


The Born matrix elements have been
computed using the helicity-amplitude technique of
refs.~\cite{Hagiwara:1985yu,Hagiwara:1988pp}. By keeping the amplitude
uncontracted with respect to the gluon polarization vector, the
spin-correlated Born matrix elements ($\mathcal{B}_{\mu\nu}$, defined
in eq.~(2.8) of ref.~\cite{Frixione:2007vw}) are easily obtained.
The color-connected squared amplitudes ($\mathcal{B}_{ij}$, defined in
eq.~(2.97) of ref.~\cite{Frixione:2007vw}) for this process are not
proportional to the Born squared amplitudes. They have been computed
inserting the color factors correspondent to the product of two color
operators ${\bf T}_i\mydot{\bf T}_j$ while computing the terms
needed to build the Born squared matrix elements.\footnote{Similarly,
  the relative weights of each colour structure present in the Born
  processes, in the limit of large number of colours, have been
  computed combining the terms
  needed to build the Born matrix elements. These weights are used to
  probabilistically assign a planar colour structure to the
  underlying-Born kinematics, from which a color flow is attached to
  the generated events, using the prescription described in
  refs.~\cite{Frixione:2007vw,Alioli:2010xd}.}  Real corrections have
been obtained using MadGraph4~\cite{Alwall:2007st}, whereas the
virtual corrections, first computed in ref.~\cite{Bern:1997sc}, have
been obtained linking the \BOX{} with
\Blackhat{}~\cite{Berger:2008sj}, using the interface proposed in
ref.~\cite{Binoth:2010xt}. A sample of Feynman diagrams that enter the
Born and the virtual contributions is reported in
fig.~\ref{fig:zjj_b_v}, whereas some numerical values for the finite
part of the virtual corrections are reported in Appendix~\ref{app:virtuals}.

\begin{figure}[htb]
  \begin{center}
    \epsfig{file=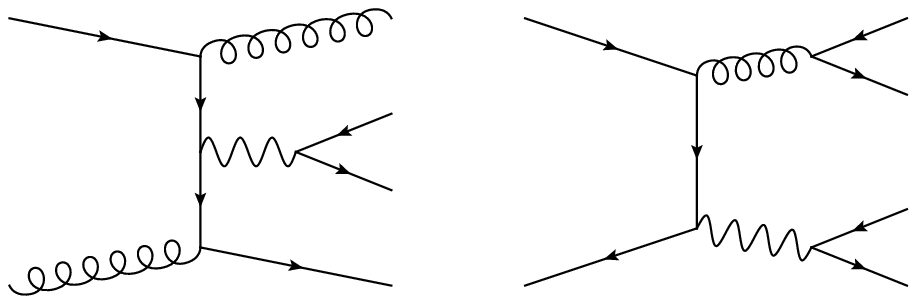,width=0.8\textwidth}\\
    ~\\
    \epsfig{file=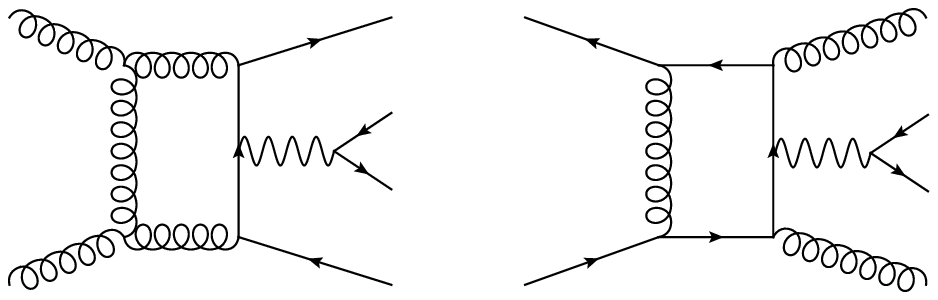,width=0.8\textwidth}\\
  \end{center}
  \caption{\label{fig:zjj_b_v} Sample graphs for the Born and the
    virtual contributions to the $Z/\gamma +2$ jets production
    process.}
\end{figure}

Several checks have been performed at this stage: spin-correlated and
color-linked squared amplitudes have been checked numerically, by
comparing the soft and collinear limits of the real matrix elements
with the expected value obtained from the factorization formulas for
collinear and soft emissions, respectively.
Thanks to the aforementioned interface, \Blackhat{} also returns the
coefficients of the double and the single pole of the one-loop
corrections: this allows for an extra cross check of the color-linked
amplitudes, or, viceversa, a check that the interface for the 2 codes
is working properly.  As a final important check, several
distributions have also been compared against NLO predictions obtained
from n-tuples generated with
\Blackhat{}+\SHERPA{}~\cite{Gleisberg:2008ta,Krauss:2001iv}, and
agreement has been found.

The significant property of a process like $V+2$ jets, that makes it
more difficult to implement than $V+1$ jet, is the fact that the Born
matrix elements are singular in several regions.  In order to suppress
the singularities of the underlying-Born amplitudes, instead of using
sharp generation cuts, we used a suppression factor that vanishes when
at least one of these regions is approached.  As explained
in~\cite{Alioli:2010qp}, this means that the underlying-Born
kinematics is generated according to a modified $\bar{B}$ function
\begin{equation}
  \label{eq:bbar}
  {\bar{B}}_{\rm supp}(\PSn) = \bar{B}(\PSn)\ F(\PSn)\,,
\end{equation}
where $\bar{B}$ is the inclusive NLO cross section at fixed
underlying-Born variables, and, for the case at hand, we used
\begin{equation}
\label{eq:supp}
F(\PSn)=
\(\frac{p_{T,1}^2}{p_{T,1}^2 +\Lambda_{\pt}^2}\)^{k_{\sss\rm IS}}
\(\frac{p_{T,2}^2}{p_{T,2}^2 +\Lambda_{\pt}^2}\)^{k_{\sss\rm IS}}
\(\frac{s_{1,2}}{s_{1,2} +\Lambda_{m}^2}\)^{k_{\sss\rm FS}}\,.
\end{equation}
$\PSn$ is the Born phase space, $p_{T,1}$ and $p_{T,2}$ are the
transverse momenta of the 2 outgoing light partons in the
underlying-Born kinematics and $s_{1,2}$ is the invariant mass squared
obtained from their momenta. Since the Born-like kinematics, from
which the radiation is generated, is distributed according to
eq.~(\ref{eq:bbar}), the generated events will be assigned a weight
proportional to $1/F$. To obtain the results shown in the next
section, we have set ${k_{\sss\rm IS}}={k_{\sss\rm FS}}=2$,
$\Lambda_{\pt}=10$ GeV and $\Lambda_{m}=5$ GeV. We also notice here
that a similar method has been recently used in
ref.~\cite{Campbell:2012am} to implement in \POWHEG{} the $H+2$ jets
process.

\section{Results}
\subsection{Comparison with NLO predictions}
\label{sec:POW_vs_NLO}
In this section we present results obtained after showering and
hadronizing the partonic events generated with \POWHEG{}. We have used
\PYTHIA{}~6.4.25 with the ``Perugia 0'' tune enabled.  We have
considered $Z(\to e^+e^-)+2$ jets production at the LHC, with an
hadronic center-of-mass energy $\sqrt{S}=7 $ TeV.

The values of physical parameters entering the computation are the following:
\begin{equation}
 m_Z=91.1876 \ \mbox{GeV}\,, \ \ \ \Gamma_Z=2.49 \ \mbox{GeV}\,, \ \ \ \ \alpha_{\rm em}^{-1}=128.802\,, \ \ \ \ \sin^2\theta_W=0.23\,,
\end{equation}
and we have used the CTEQ6M~\cite{Pumplin:2002vw} parton distribution
functions. In the computation of the $\bar{B}$ function, the
renormalization and factorization scales have been chosen equal to
$\hat{H}_T/2$, where
\begin{equation}
  \hat{H}_T=\sqrt{m_Z^2+p_{T,Z}^2}+p_{T,1}+p_{T,2}\,,
\end{equation}
$p_{T,Z}$ is the transverse momentum of the $Z$-boson, and all the
quantities are computed using the underlying-Born kinematics.

As far as technical parameters or special options of the \BOX{}
program are concerned, we have kept them equal to their default value,
and we have not used any folding in the integration of the radiation
variables~\cite{Alioli:2010qp,Alioli:2010xd}.  In so doing, the
fraction of negative-weight events amounts to be $21\%$. 
These events were kept in the final sample used for the analysis,
although of course not all of them will contribute to the plots, since
some will not pass the cuts.

Jets have been defined according to the anti-$\kt$
algorithm~\cite{Cacciari:2008gp,Cacciari:2005hq,Cacciari:2011ma},
setting $R=0.4$, and the following cuts have been enforced in the
analysis:
\begin{eqnarray}
\label{eq:cuts}
&~&66.328\ \mbox{GeV}<m_{e^+ e^-}< 116.048\ \mbox{GeV}\,, \ \ \ p_{T,e}>20\ \mbox{GeV}\,, \ \ \ |y_{e}|<2.5\,, \nn\\
&~&p_{T,j}>30\ \mbox{GeV}\,,\ \ \ |\eta_{j}|<4.4\,.
\end{eqnarray}
Events are accepted only if there are at least 2 jets passing the
above cuts. We also remind that we have not used isolation cuts for
the leptons, and that photon radiation off leptons have been switched
off in the shower,\footnote{Photon radiation has been switched off by
  setting {\ttfamily MSTJ(41)=3} in \PYTHIA{}.} in order to allow for
a clear comparison with the fixed order result.

\begin{figure}[!htb]
  \begin{center}
    \epsfig{file=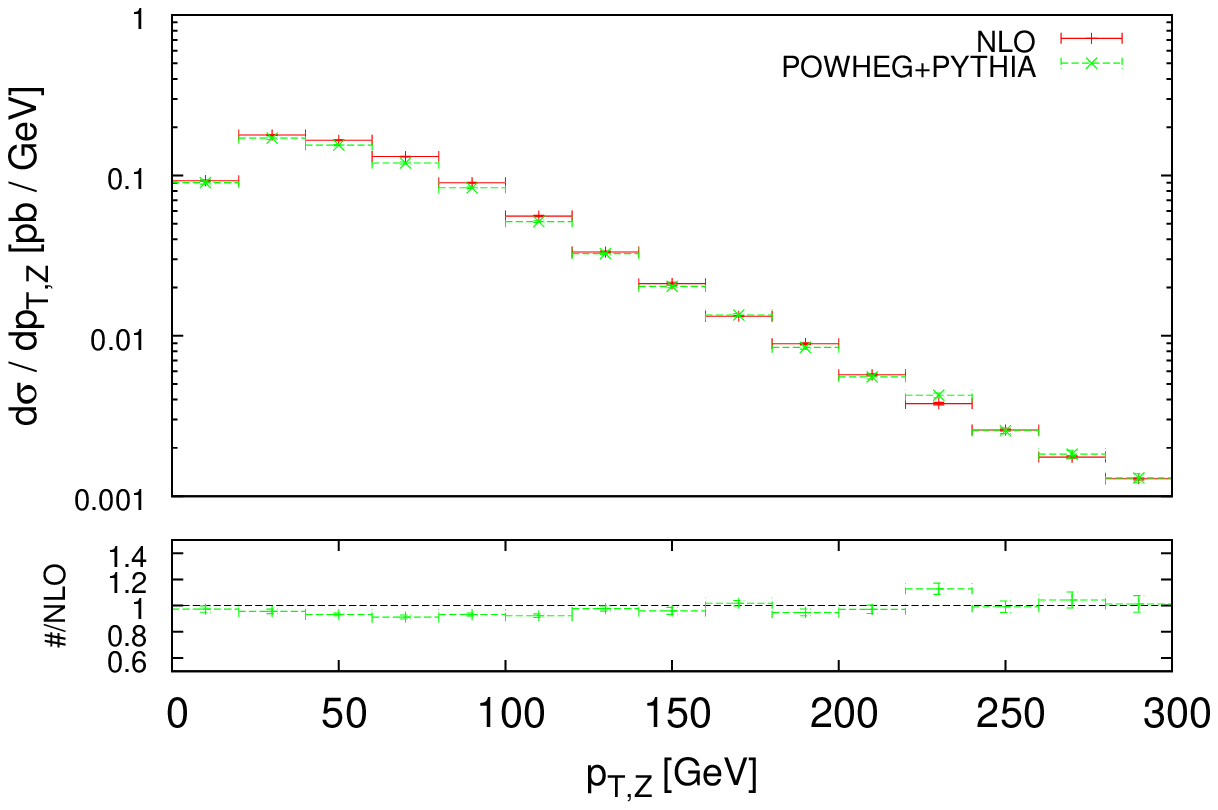,width=0.5\textwidth}~
    \epsfig{file=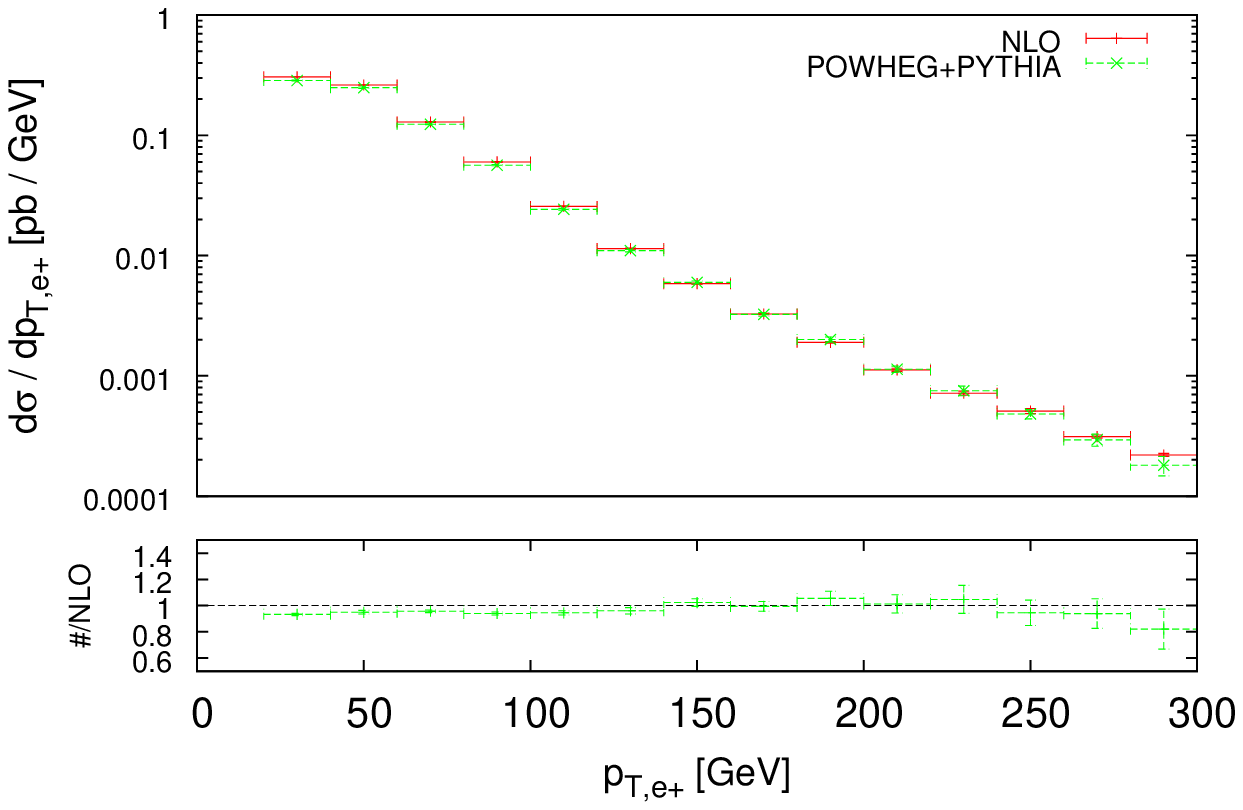,width=0.5\textwidth}\\
    \epsfig{file=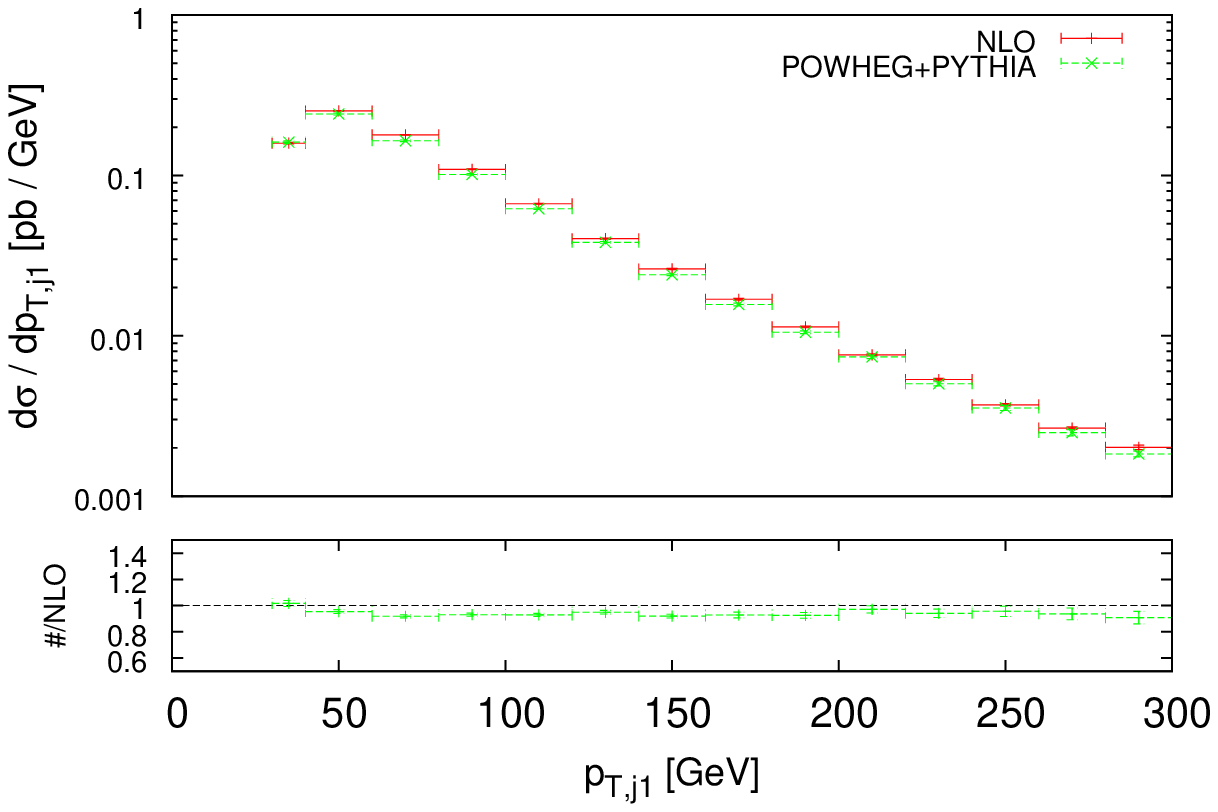,width=0.5\textwidth}~
    \epsfig{file=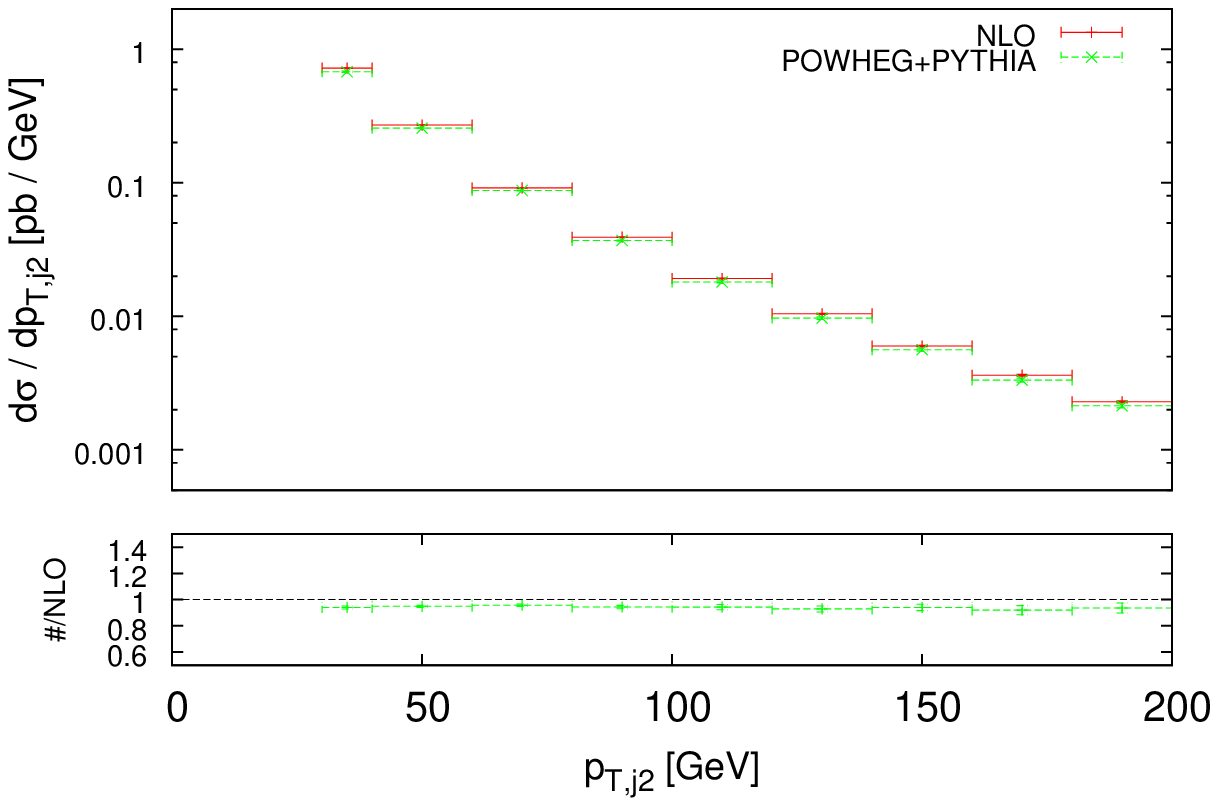,width=0.5\textwidth}
  \end{center}
  \caption{\label{fig:ptZpt1pt2} Comparisons between \POWHEG{}
    (interfaced to \PYTHIA{}) and the NLO results at the LHC $pp$
    collider ($\sqrt{S}=7$ TeV), for the transverse momenta of the
    reconstructed $Z$-boson, of the positron (upper panel), of the
    hardest and of the second-hardest jet (lower panel). Vertical bars
    correspond to statistical errors.}
\end{figure}

\begin{figure}[!htb]
  \begin{center}
    \epsfig{file=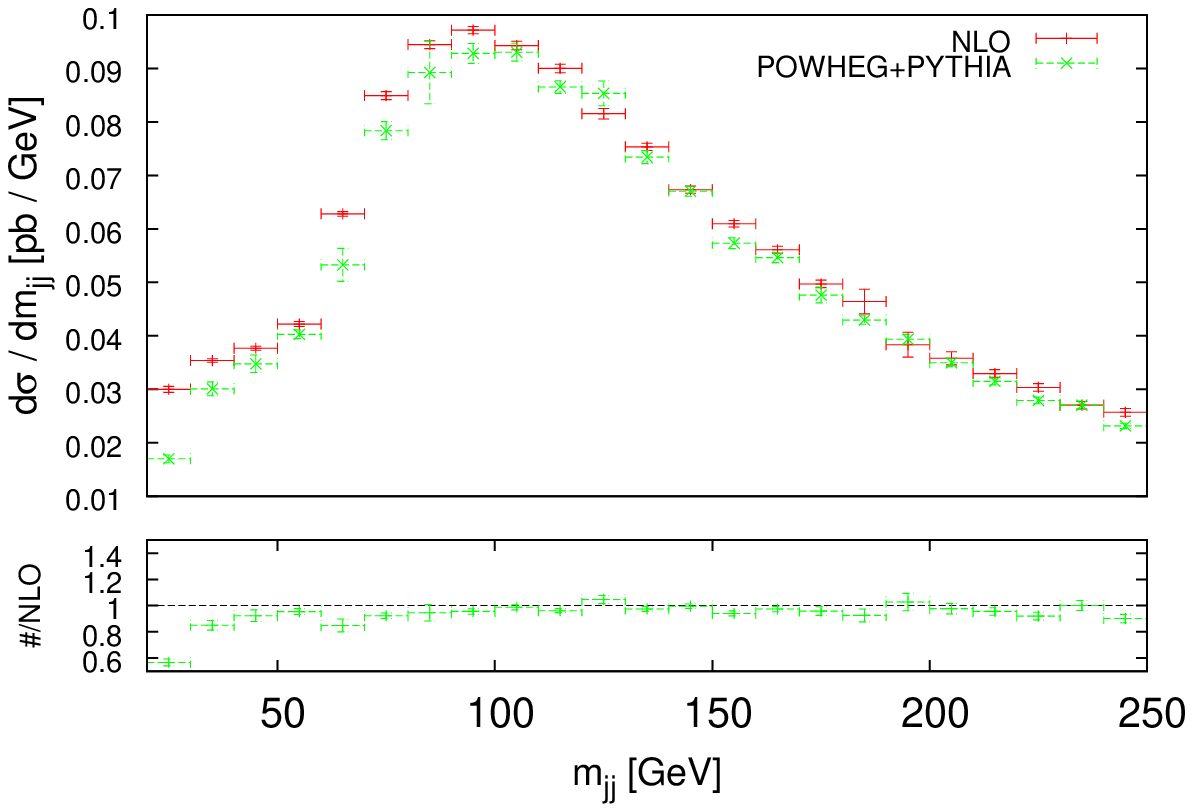,width=0.5\textwidth}~
    \epsfig{file=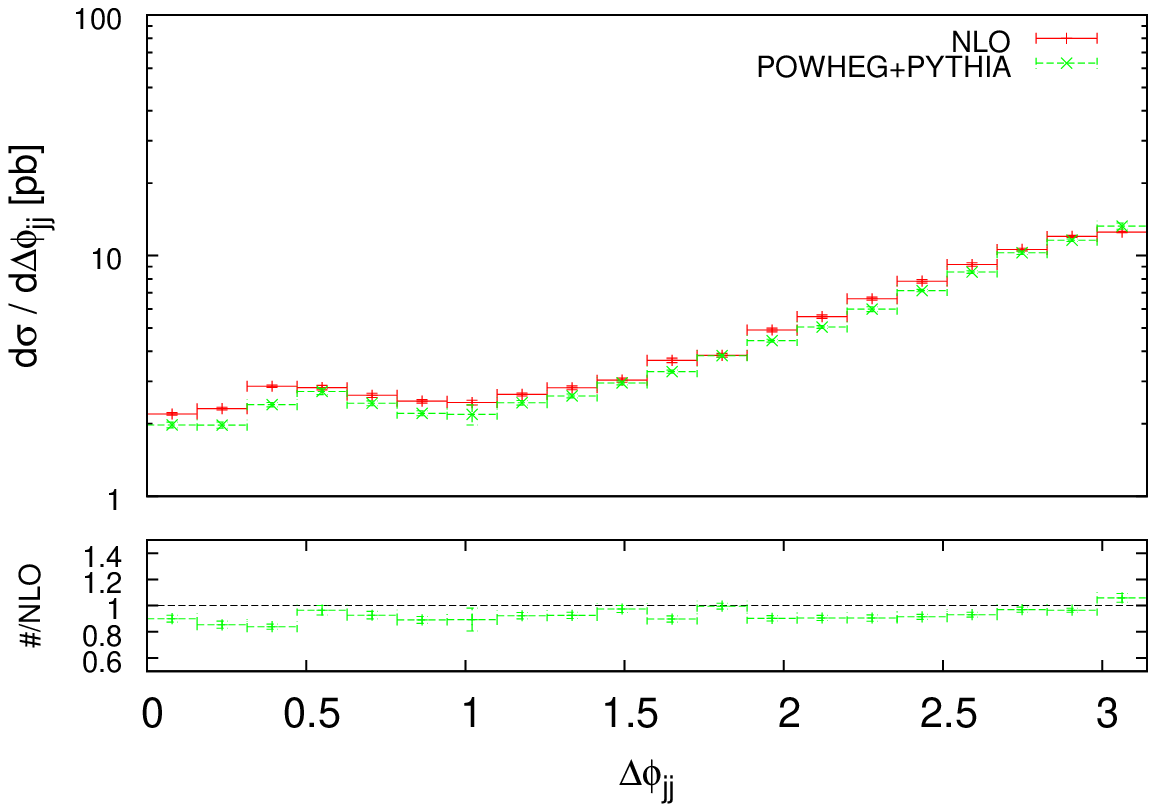,width=0.5\textwidth}
  \end{center}
  \caption{\label{fig:m12dphi12Ht} Comparisons between \POWHEG{}
    (interfaced to \PYTHIA{}) and the NLO results at the LHC $pp$
    collider ($\sqrt{S}=7$ TeV), for the invariant mass of the two
    hardest jets and for their azimuthal separation. Vertical bars
    correspond to statistical errors.}
\end{figure}

In fig.~\ref{fig:ptZpt1pt2} we show the transverse momentum spectra of
the reconstructed $Z$ boson ($p_{T,Z}$), of the positron
($p_{T,e^+}$), and of the hardest and second-hardest jet
($p_{T,j_1},p_{T,j_2}$), whereas in fig.~\ref{fig:m12dphi12Ht} the
invariant mass of the system made by the 2 hardest jets ($m_{jj}$) and
their azimuthal separation ($\Delta\Phi_{jj}$) is shown.

The agreement between the NLO predictions and the results after the
shower and the hadronization stage is very good for these inclusive
observables. Shapes are essentially unmodified, and very small
differences can be observed in the lower panel of each plot, where the
ratio between the showered result and the NLO is
plotted.\footnote{Errors in the lower panels of all the plots in this
  article represent the error on the ratio between two results,
  obtained propagating the errors of each set.}
We have checked that they are of the same order of the
theoretical uncertainty of the NLO result, obtained by changing the
renormalization and factorization scales by the usual factors
$\{1/2,2\}$. It is also worth noticing that effects of the same order
have been observed in other NLO+PS implementations of similar
complexity~\cite{Frederix:2011ig,Hoeche:2012ft}, although in presence
of different cuts.

\begin{figure}[!htb]
  \begin{center}
    \epsfig{file=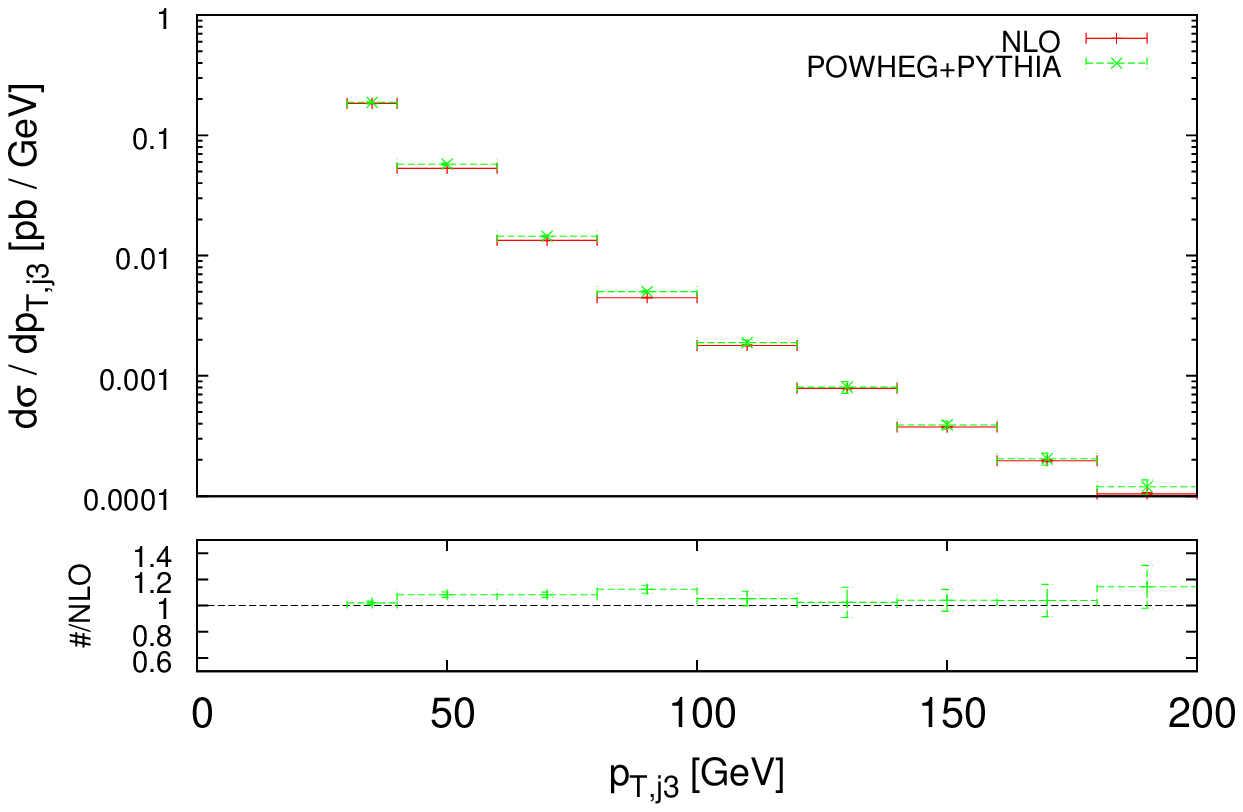,width=0.5\textwidth}~
    \epsfig{file=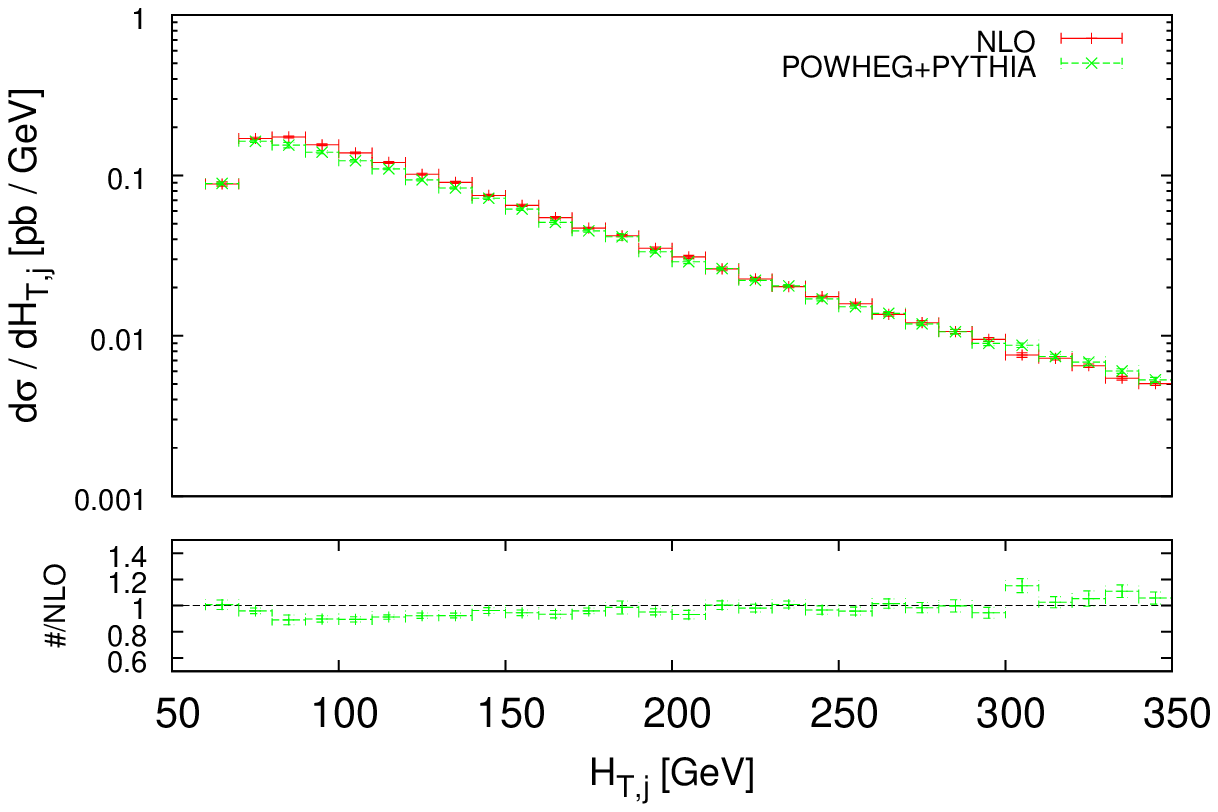,width=0.5\textwidth}\\
    \epsfig{file=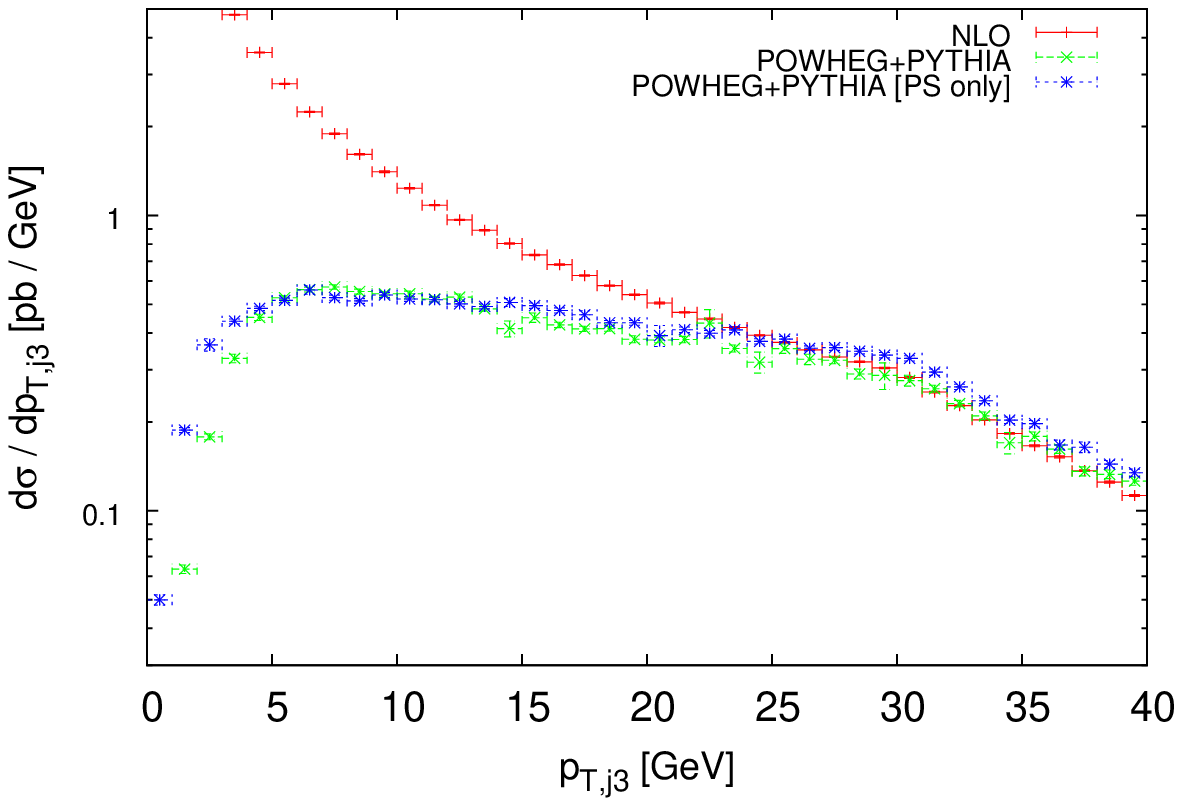,width=0.5\textwidth}~
  \end{center}
  \caption{\label{fig:pt3pt3Ht} Comparisons between \POWHEG{}
    (interfaced to \PYTHIA{}) and the NLO results at the LHC $pp$
    collider ($\sqrt{S}=7$ TeV), for the third hardest jets
    transverse momentum and for $H_{T,j}$ (upper panel). In the lower
    panel the low cut on the jets transverse momenta is kept only for
    the hardest and second-hardest jet. Vertical bars correspond to
    statistical errors.}
\end{figure}

In the upper panel of fig.~\ref{fig:pt3pt3Ht} we show instead the
transverse momentum of the third hardest jet and $H_{T,j}$, which is
defined to be the scalar sum of all the jets transverse momenta: good
agreement is found between the NLO result and the \POWHEG{}+\PYTHIA{}
prediction for these observables as well. In particular, for values of
$p_{T,j_3}$ far from the low $\pt$ region, the results after the
showering and hadronization stage agree with the fixed order result
(which has leading-order accuracy for this observable).  In the small
transverse momentum region, the NLO+PS prediction for $p_{T,j_3}$ is
expected to show Sudakov damping, as opposite to the NLO result, which
diverges: this is noticeable in the lower panel of
fig.~\ref{fig:pt3pt3Ht}, where we plot $p_{T,j_3}$ removing the lower
cut on the transverse momentum of all but the two hardest jets. The
results obtained just after the parton-shower stage performed by
\PYTHIA{} are also shown for this observable. Sudakov damping is
clearly present in all the curves where resummation of soft-collinear
emissions is performed, although in the very low-$\pt$ region
non-perturbative effects seem to play a small but noticeable role.

We have also performed the same analysis using as factorization and
renormalization scale the $Z$-boson transverse mass, computed with the
underlying-Born kinematics ($\mu=E_{T,Z}=\sqrt{m_Z^2+p_{T,Z}^2}$), and
we have found similar results.

As mentioned in the introduction, $Zjj$ production is a background for
Higgs-boson production via VBF, when the $H\to\tau\tau$ decay channel
is considered in the analysis. For this reason it is interesting to
check how typical distributions look like for the QCD $Zjj$
background, in presence of a minimal set of VBF cuts. We chose
\begin{eqnarray}
  \label{eq:cutsVBF}
  &~&p_{T,\ell}>20\ \mbox{GeV}\,, \ \ \ |y_{\ell}|<2.5\,, \nn\\
  &~&|\eta_{j}|<5.0\,, \ \ \ p_{T,j}>20\ \mbox{GeV}\,, \ \ \ p_{T,j_{\rm tag}}>30\ \mbox{GeV}\,, \nn\\
  &~&|\eta_{j_1}-\eta_{j_2}|>4.0\,, \ \ \ \eta_{j_1}\cdot\eta_{j_2}<0\,, \nn\\ 
  &~&\min{(\eta_{j_1},\eta_{j_2})} + 0.4 < \eta_{\ell^+/\ell^-} < \max{(\eta_{j_1},\eta_{j_2})} - 0.4\,,
\end{eqnarray}
and we restricted the invariant mass of the lepton pair to lie in the
interval $[66.328,116.048]$ GeV.

The ``tagging jets'' are the two hardest jets, and we have used as
before the anti-$\kt$ algorithm, with $R=0.4$. For this analysis we
have generated the events using as factorization and renormalization
scale the $Z$-boson transverse mass, computed with the underlying-Born
kinematics. Moreover, multiple parton interactions have been switched
off. 

\begin{figure}[!htb]
  \begin{center}
    \epsfig{file=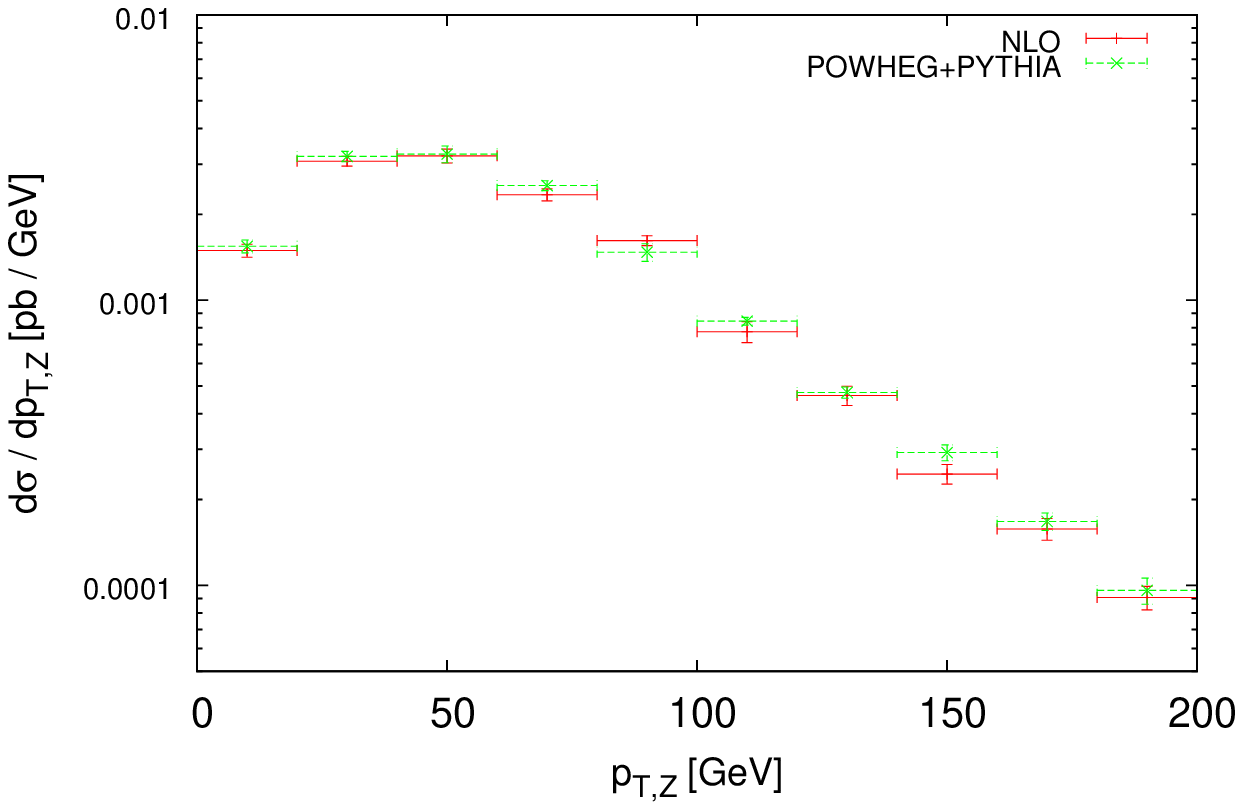,width=0.5\textwidth}~
    \epsfig{file=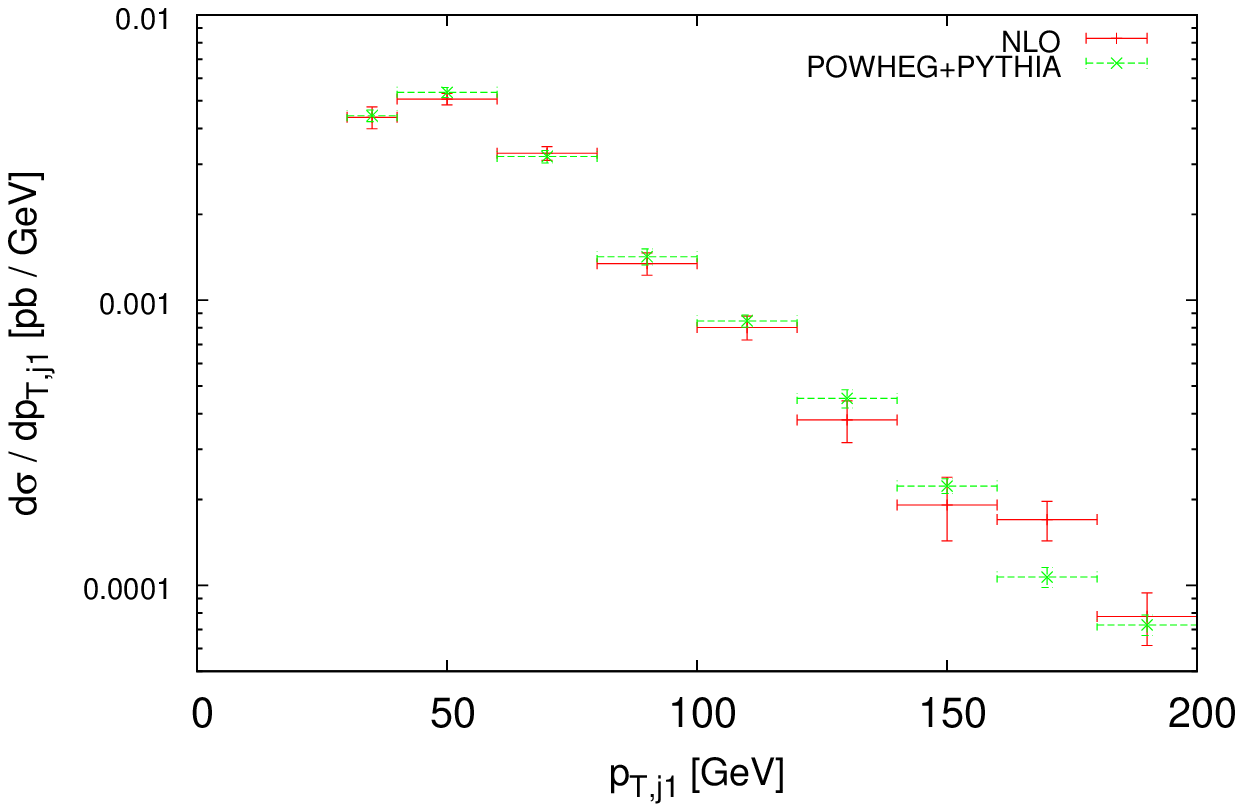,width=0.5\textwidth}\\
    \epsfig{file=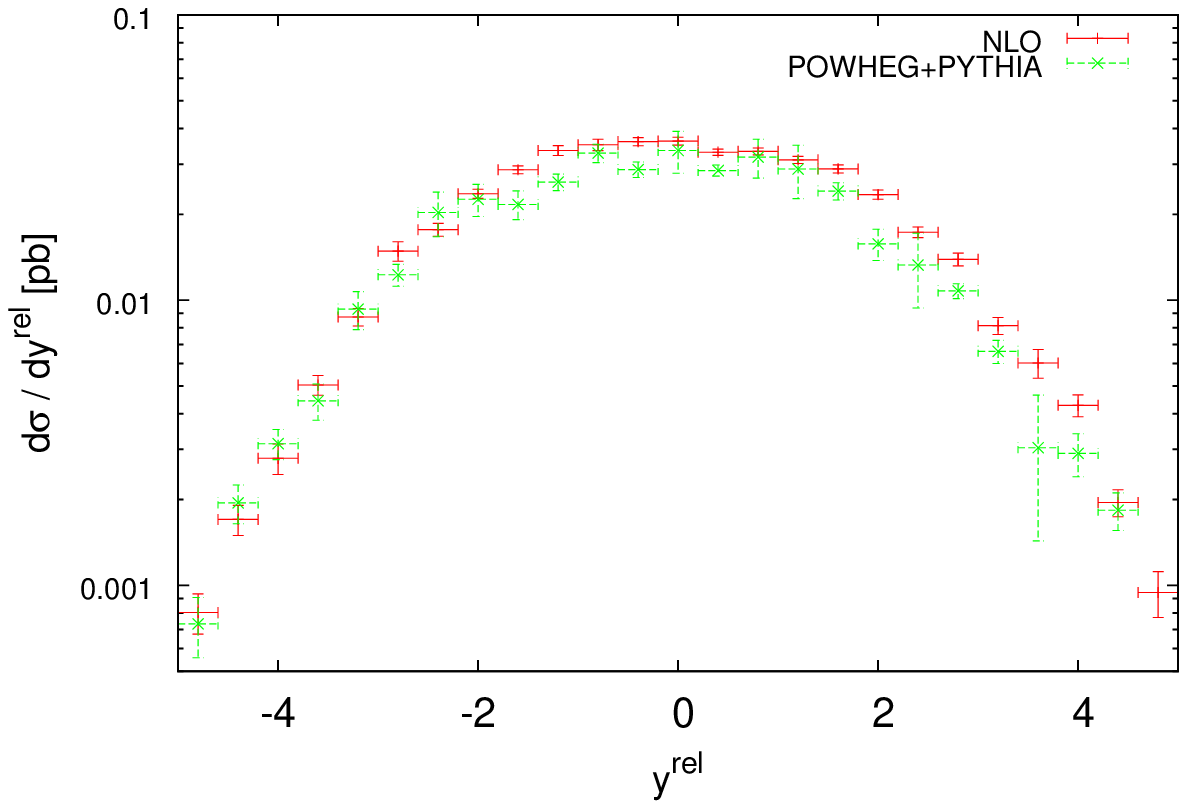,width=0.5\textwidth}~
    \epsfig{file=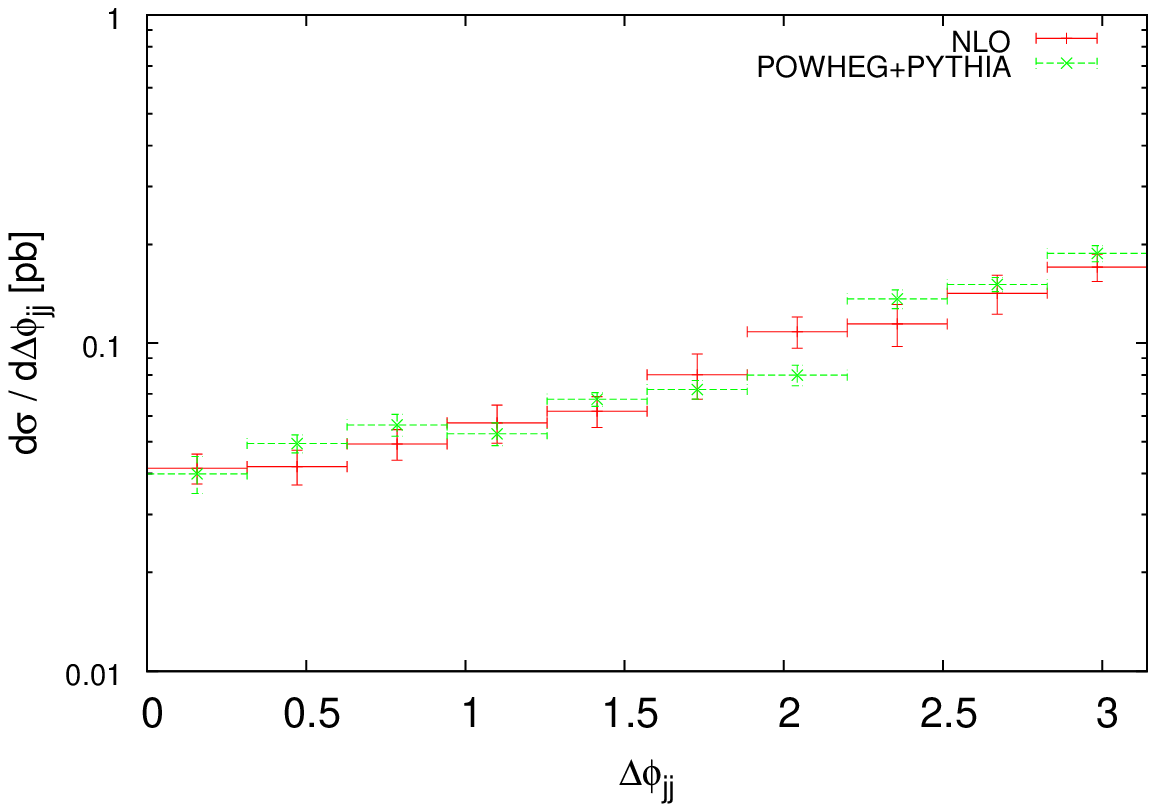,width=0.5\textwidth}
  \end{center}
  \caption{\label{fig:vbf} Comparisons between \POWHEG{} (interfaced
    to \PYTHIA{}) and the NLO results at the LHC $pp$ collider
    ($\sqrt{S}=7$ TeV) in presence of the VBF cuts of
    eq.~(\ref{eq:cutsVBF}): in the upper panel the transverse momenta
    of the reconstructed $Z$-boson and of the hardest jet are shown,
    whereas in the lower panel the azimuthal decorrelation between the
    tagging jets and $y^{\rm rel}=y_{j_3}-(y_{j_1}+y_{j_2})/2$ are
    reported. Vertical bars correspond to statistical errors.}
\end{figure}

In the upper panel of fig.~\ref{fig:vbf} we show $p_{T,Z}$ and
$p_{T,j_1}$, whereas in the lower panel we show the azimuthal
decorrelation between the two hardest jets and the shifted rapidity
$y^{\rm rel}=y_{j_3}-(y_{j_1}+y_{j_2})/2$, which is an useful quantity
to measure the distance between the tagging jets and the third hardest
jet.

Since we are using cuts designed to suppress the $Zjj$ background, the
statistical significance of the plots in fig.~\ref{fig:vbf} is not
optimal. However, we found that the results are not changed sizeably
when going from a NLO to a NLO+PS prediction.  In particular $y^{\rm
  rel}$ is peaked at 0, whereas, as a consequence of the VBF cuts, the
rapidity of the two tagging jets (not shown) is peaked at $|y_{j_{\rm
    tag}}|\simeq 2.5$. This is the expected behaviour, since the QCD
production of $Zjj$ is a process dominated by quark or gluon exchange
in the $t$-channel, and therefore one expects to observe a third jet
in the central rapidity region, as opposite to what happens in
processes dominated by a $t$-channel exchange of a colourless state,
like Higgs-boson production via VBF or $Zjj$ EW production. In the
latter cases, the extra jet activity tends to be close to one of the
two tagging jets, and, as a consequence, $y^{\rm rel}$ peaks at high
rapidities and exhibits a dip at $y^{\rm
  rel}=0$~\cite{Rainwater:1996ud,Nason:2009ai}.  The azimuthal
decorrelation between the tagging jets is also particularly relevant:
indeed it was shown that properties of the Higgs-boson such as its
parity could be measured when it is produced via VBF, by looking at
the azimuthal separation between the tagging
jets~\cite{Plehn:2001nj}. This distribution appears to be stable as
well for the $Zjj$ background, when going from NLO to NLO+PS accuracy.

\subsection{Comparison with LHC data}
\label{sec:POW_vs_data}
The ATLAS Collaboration has published a study on the production of a
$Z$-boson in association with jets~\cite{Aad:2011qv}, and good
agreement has been found when comparing the experimental results with
QCD NLO perturbative predictions and predictions from Monte Carlo
generators implementing LO matrix elements matched with parton
showers.  In this section we show a comparison of our NLO+PS
predictions with ATLAS data.  The set of cuts we used is very similar
to that of eq.~(\ref{eq:cuts}). Here we ask for
\begin{eqnarray}
\label{eq:cutsATLAS}
&~&66\ \mbox{GeV}<m_{e^+ e^-}< 116\ \mbox{GeV}\,,\nn\\
&~&p_{T,e}>20\ \mbox{GeV}\,, \ \ \ |y_{e}|<2.5\,, \ \ \ \Delta R_{j,e}>0.5\,,\nn\\
&~&p_{T,j}>30\ \mbox{GeV}\,,\ \ \ |y_{j}|<4.4\,,
\end{eqnarray}
and jets are built with the anti-$\kt$ algorithm, with $R=0.4$.

\begin{figure}[!htb]
  \begin{center}
    \epsfig{file=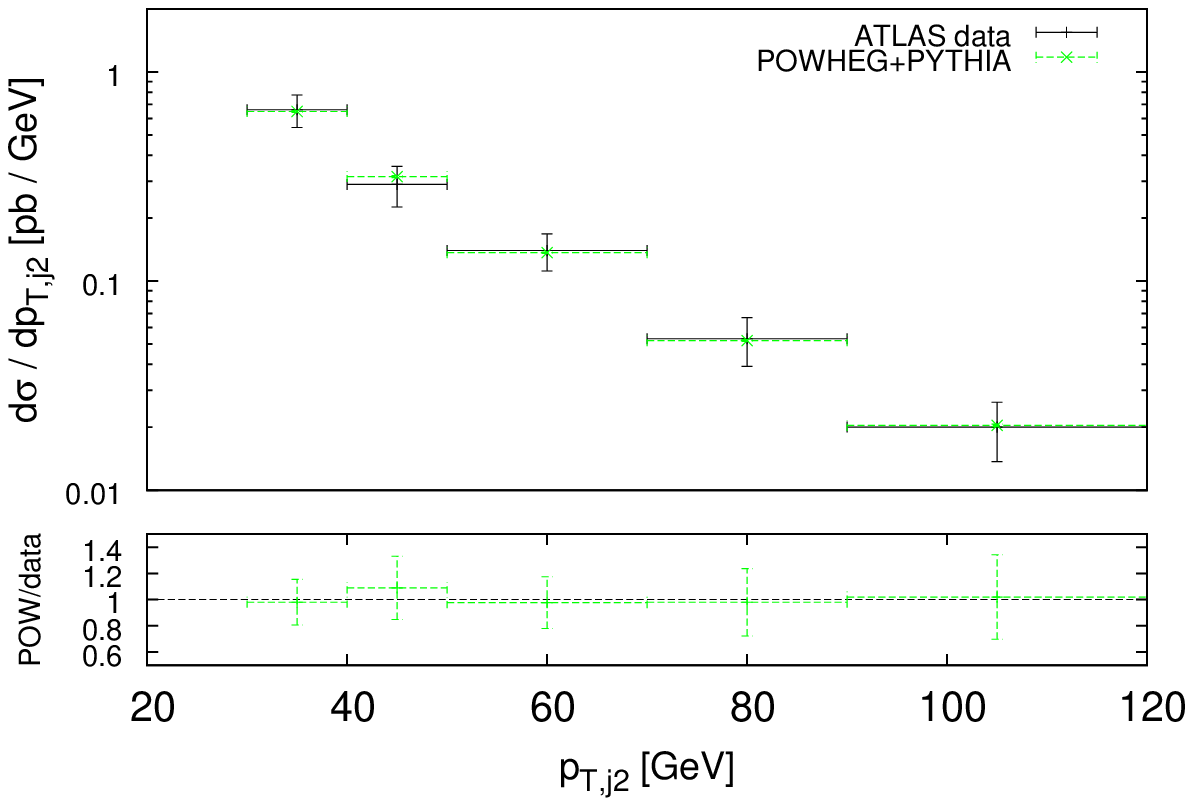,width=0.5\textwidth}~
    \epsfig{file=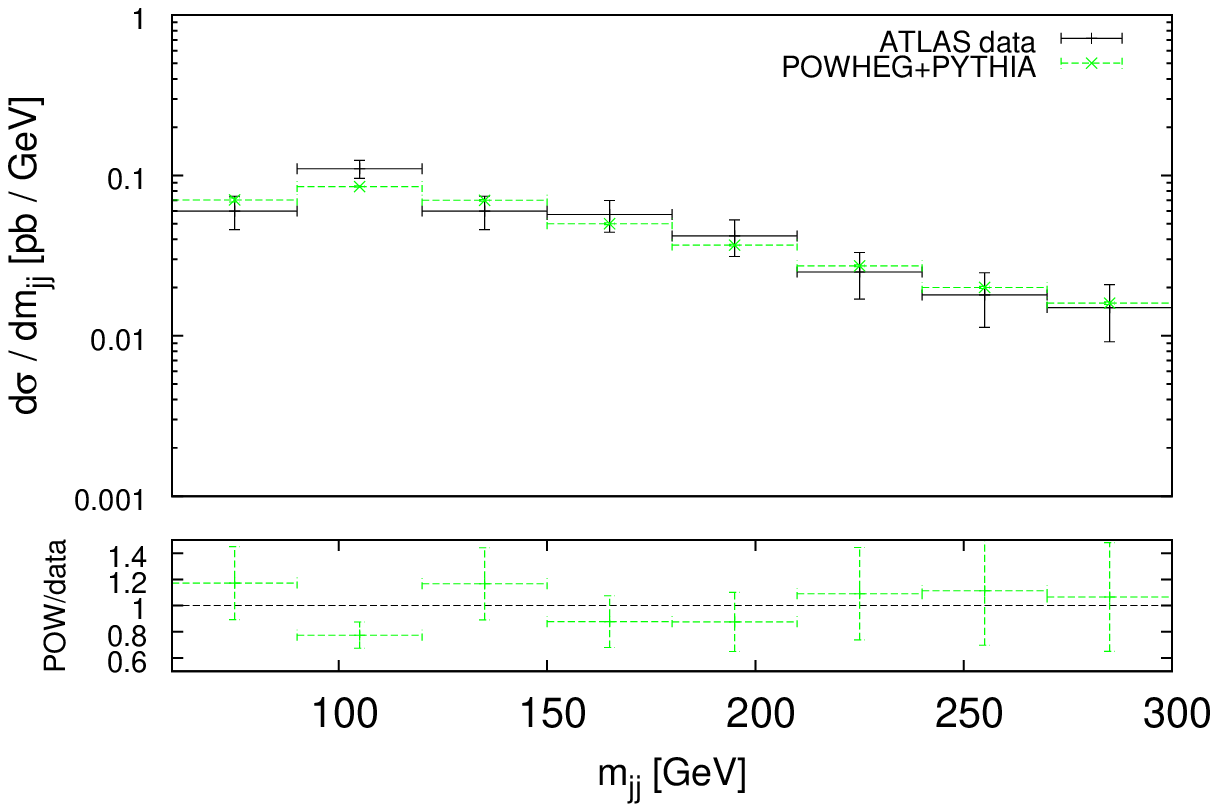,width=0.5\textwidth}\\
    \epsfig{file=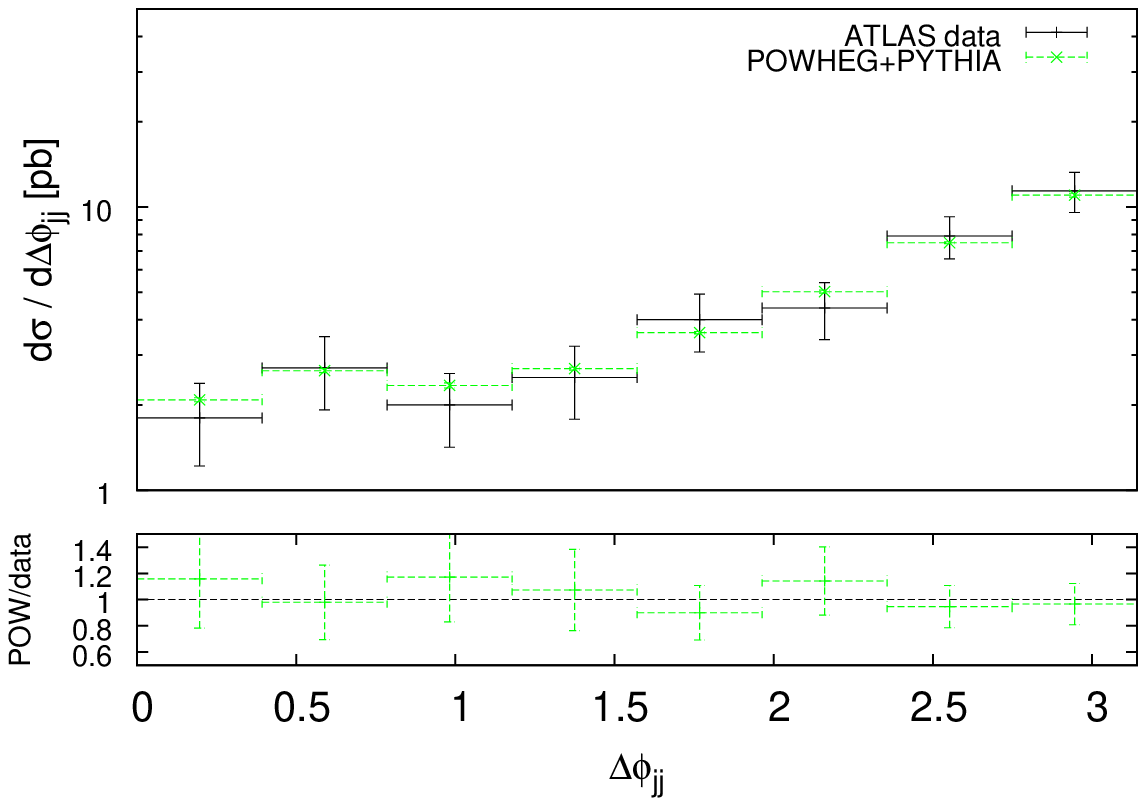,width=0.5\textwidth}~
    \epsfig{file=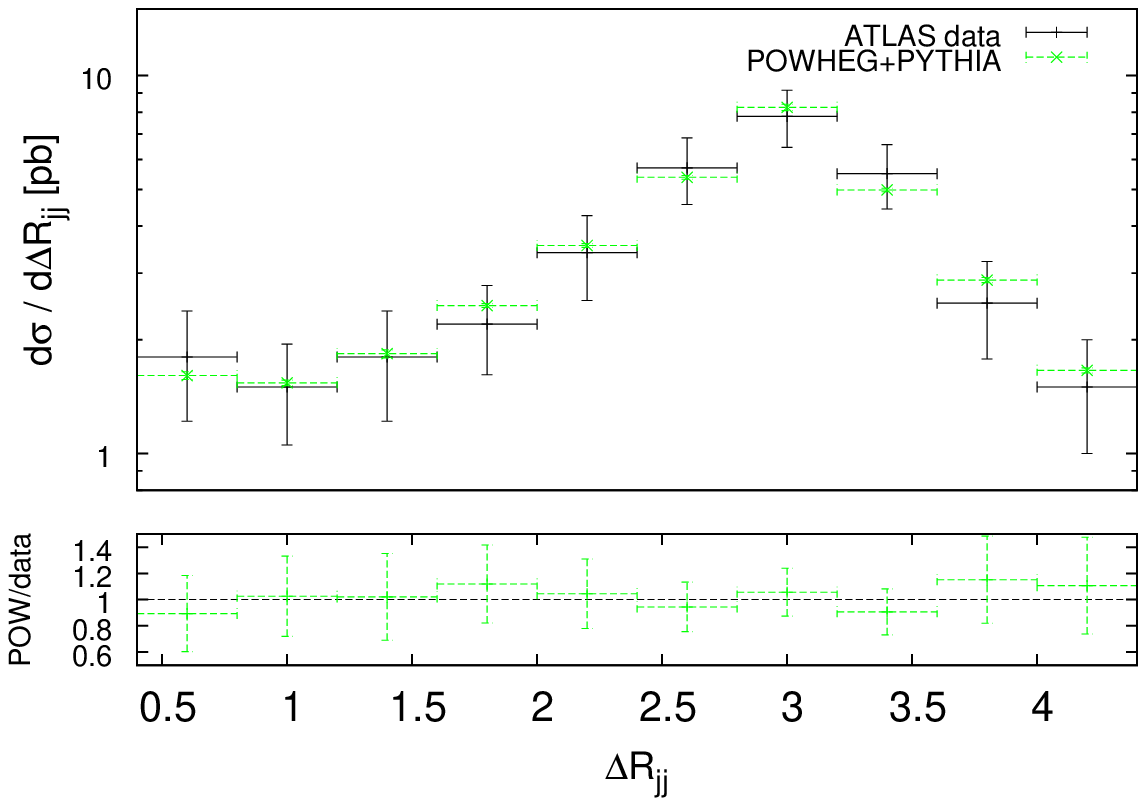,width=0.5\textwidth}
  \end{center}
  \caption{\label{fig:ATLAS} Comparisons between \POWHEG{} (interfaced
    to \PYTHIA{}) and the ATLAS data ($\sqrt{S}=7$ TeV), in presence
    of the cuts of eq.~(\ref{eq:cutsATLAS}), for the transverse
    momentum of the second-hardest jet, the invariant mass of the two
    hardest jets (upper panel), the azimuthal separation and the
    $\Delta R$ distance between the two hardest jets (lower panel).
    Vertical bars on the \POWHEG{} results correspond to statistical
    errors, whereas bars on ATLAS data represent the sum (in
    quadrature) of the statistical and systematical errors reported in
    ref.~\cite{Aad:2011qv}.}
\end{figure}

In fig.~\ref{fig:ATLAS} we show our predictions, obtained with the
sample generated with $\mu=\hat{H}_T/2$, together with the ATLAS data.
In the upper panel we show the transverse momentum of the
second-hardest jet and the invariant mass of the two hardest jets, for
events with at least 2 jets. In the lower panel instead, the azimuthal
separation $\Delta\Phi_{jj}$ between the two hardest jets and their
distance $\Delta
R_{jj}=\sqrt{\(\Phi_{j_1}-\Phi_{j_2}\)^2+\(y_{j_1}-y_{j_2}\)^2}$ in the
$\phi-y$ plane is reported. The outcome of these comparisons is that
the agreement between experimental data and the \POWHEG{} results is
very good.

\section{Conclusions}

In this paper we have described the first implementation of $Z+2$ jets
production at next-to-leading order in QCD, in the \POWHEG{}
framework. We have used the \BOX{} package, which is a program that
automates the algorithm first proposed in ref.~\cite{Nason:2004rx} and
then described in detail in ref.~\cite{Frixione:2007vw}.  The main
purpose of this paper was to show that no particular problems occur in
the implementation of complicated processes with \POWHEG{}.  This has
been recently observed in ref.~\cite{Campbell:2012am}, and we obtained
similar findings. We regulated the divergent underlying-Born process
by using a damped $\bar{B}$ function, and we have shown that results
are in good agreement with the fixed order for observables where they
are expected to. Sudakov damping is instead present for observables
sensitive to multiple soft-collinear emissions, such as the low-$\pt$
region of the transverse momentum of the third hardest jet.

An important phenomenological application which can be performed
having this processes available with NLO+PS accuracy is a study of the
QCD radiation patterns of $Zjj$ with respect to those present in $Hjj$
production, especially in presence of cuts used in VBF
searches. 
We have given an example of such a study by checking that typical
observables in $Zjj$ production in presence of VBF cuts have the
behaviour expected from general properties of QCD, and they are not
affected sizeably when going from NLO to a NLO+PS description.
Further studies along these lines could also be performed using NLO+PS
implementations of $Hjj$ via gluon~\cite{Campbell:2012am} and
vector-boson fusion~\cite{Nason:2009ai,D'Errico:2011um} and
$t\bar{t}$(+ jets)~\cite{Frixione:2007nw,Kardos:2011qa,Alioli:2011as},
together with the $Zjj$ implementation presented here.

Moreover, it could also be interesting to merge NLO+PS predictions for
$Z+2$ jets together with the same predictions for lower
multiplicities, using for example the approach proposed in
ref.~\cite{Alioli:2011nr}. These studies are left for future work.

We have also compared our NLO+PS results with recent ATLAS data, and
we have found good agreement.

The computer code for this \POWHEG{} implementation will soon be
available within the public branch of the \BOX{} package. It will be
possible to link it with public one-loop codes
to evaluate the virtual corrections. 


\acknowledgments
I would like to thank D.~Ma\^{\i}tre for providing me with a version
of the \Blackhat{} code, and for very useful informatic help.

\appendix
\section{One-loop virtual amplitudes}
\label{app:virtuals}
In this appendix, we report the numerical values of the finite part of
virtual corrections for the following phase space point:
\begin{eqnarray}
\label{eq:momenta}
p_\splus  &=&  ( 138.4784456174 ;\   0.0000000000 ,\   0.0000000000 ,\  138.4784456174 )\ \mbox{GeV}\,, \nn\\
p_\sminus &=&  ( 138.4784456174 ;\   0.0000000000 ,\   0.0000000000 ,\ -138.4784456174 )\ \mbox{GeV}\,, \nn\\
p_{e^-} &=&    (  39.8695145736 ;\ -30.2118421708 ,\   7.3821554172 ,\ -24.9464740270  )\ \mbox{GeV}\,, \nn\\
p_{e^+} &=&    (  61.7949044527 ;\  60.7995170788 ,\  -7.3821554172 ,\  -8.2178294395  )\ \mbox{GeV}\,, \nn\\
p_1 &=&       ( 104.9831601822 ;\  -38.2721470356 ,\  44.5354030781 ,\ 87.0247353101  )\ \mbox{GeV}\,, \nn\\
p_2 &=&        ( 70.3093120261 ;\    7.6844721276 ,\ -44.5354030781 ,\ -53.8604318436 )\ \mbox{GeV}\,. \nn\\
\end{eqnarray}

The finite parts of the interference between the tree-level and the
one-loop amplitude (as returned by \Blackhat{}), computed in the 't
Hooft-Veltman scheme and stripped off of the usual factor
\begin{equation}
c_{\Gamma}= \frac{(4\pi)^\ep}{\Gamma(1-\ep)}\,,
\end{equation}
are reported in Table~\ref{table:virt_amplitudes}, together with the
corresponding partonic subprocesses and the Born squared amplitudes.
\begin{table}[t]
\centering
\begin{tabular}{|c|c|c|}
\hline\hline
subprocess & $ \left| \mathcal{A}^{\rm tree} \right|^2$ &
$2 \, {\Re}\{ (\mathcal{A}^{\rm tree})^* \cdot \mathcal{A}^{\rm 1-loop} \}$ \\
\hline
\hline
$u \bar{u} \to e^- e^+ g g$ & 1.0489864146 E-005 & 1.7673954550 E-004 \\
\hline
$d \bar{d} \to e^- e^+ g g$ & 1.3620189441 E-005 & 2.4684280244 E-004 \\
\hline
\hline
$u \bar{u} \to e^- e^+ u \bar{u}$ & 6.9320177545 E-006 & 2.1474734072 E-004 \\
\hline
$u \bar{u} \to e^- e^+ c \bar{c}$ & 1.8468108910 E-007 & 1.8007297663 E-006 \\
\hline
$d \bar{d} \to e^- e^+ d \bar{d}$ & 1.0023268662 E-005 & 3.1647659850 E-004 \\
\hline
$d \bar{d} \to e^- e^+ s \bar{s}$ & 2.3673784898 E-007 & 2.4179197657 E-006 \\
\hline
$u \bar{u} \to e^- e^+ d \bar{d}$ & 6.6505600059 E-007 & 1.3323314116 E-006 \\
\hline
\end{tabular}
\caption{\label{table:virt_amplitudes} Partonic subprocesses and corresponding values for the Born squared amplitudes and the interference between the tree-level and the one-loop amplitudes. Momenta are given in
  eq.~(\ref{eq:momenta}), whereas physical parameters are reported in the text. Processes that can be obtained by crossing are not reported.}
\end{table}
The physical parameters chosen are as reported in
section~\ref{sec:POW_vs_NLO}, but here the renormalization scale have
been set equal to $m_Z$. The results for the virtual amplitudes are
divided by $\as/2\pi$, \emph{i.e.} they do not contain the extra power
of the strong coupling constant.  We stress that with the \Blackhat{}
version that has been used in this work these amplitudes are computed
in the large $m_t$ limit, and the axial contributions due to fermionic
loops are not included. It has been explicitly checked that the above
results agree with the same amplitudes obtained with \MCFM{}, when the
aforementioned contributions are turned off. Agreement between
\Blackhat{} and \MCFM{} has been obtained also using $\hat{H}_T/2$ or
the $Z$-boson transverse mass as renormalization scale.

\bibliography{paper}

\end{document}